\RequirePackage{fix-cm}
\documentclass[smallextended]{svjour3}
\smartqed

\usepackage{graphicx}
\usepackage[T1]{fontenc}
\usepackage{mathptmx}
\usepackage{tikz}
\usepackage{pgfplots}
\usepackage{adjustbox}
\usepackage{amssymb}
\usepackage{amsmath}
\usepackage{mathtools}

\usepackage{color}
\usepackage{float}
\usepackage{url}

\newcommand{\el}[2]{\varepsilon^{#1}_{#2}}
\journalname{Current Pathobiology Reports}

\begin{document}

\title{Dynamics and Sensitivity of Signaling Pathways}

\author{Michael A. Kochen  \and
        Steven S. Andrews  \and
        H. Steven Wiley \and
        Song Feng \and
        Herbert M. Sauro
}

\institute{Michael A. Kochen \at
             Department of Bioengineering, University of Washington, Seattle, WA, US \\
              \email{kochenma@uw.edu}  
           \and
           Steven S. Andrews \at
           Department of Bioengineering, University of Washington, Seattle, WA, US \\
           \email{steven.s.andrews@gmail.com}
           \and
           H. Steven Wiley \at
           Environmental Molecular Sciences Laboratory, Pacific Northwest National Laboratory, Richland, WA, US \\
           \email{Steven.Wiley@pnnl.gov}
           \and
           Song Feng \at
           Biological Sciences Division, Pacific Northwest National Laboratory, Richland, WA, US \\
           \email{song.feng@pnnl.gov}
           \and
            Herbert M. Sauro at
            Department of Bioengineering, University of Washington, Seattle, WA, US \\ 
            Tel.: +206-685-2119\\
              \email{hsauro@uw.edu} 
}
\date{Received: date / Accepted: date}

\maketitle

\begin{abstract}
Purpose of Review: Signaling pathways serve to communicate information about extracellular conditions into the cell, to both the nucleus and cytoplasmic processes to control cell responses. Genetic mutations in signaling network components are frequently associated with cancer and can result in cells acquiring an ability  to  divide  and  grow  uncontrollably. Because signaling pathways play such a significant role in cancer initiation and advancement, their constituent proteins are attractive therapeutic targets. In this review, we discuss how signaling pathway modeling can assist with identifying effective drugs for treating diseases, such as cancer. An achievement that would facilitate the use of such models is their ability to identify controlling biochemical parameters in signaling pathways, such as molecular abundances and chemical reaction rates, because this would help determine effective points of attack by therapeutics. 

Recent Findings: We summarize the current state of understanding the sensitivity of phosphorylation cycles with and without sequestration. We also describe some basic properties of regulatory motifs including feedback and feed-forward regulation. 

Summary: Although much recent work has focused on understanding the dynamics and particularly the sensitivity of signaling networks in eukaryotic systems, there is still an urgent need to build more scalable models of signaling networks that can appropriately represent their complexity across different cell types and tumors.  

\end{abstract}

\section{Introduction} \label{intro}

Cellular signaling pathways serve to communicate information about extracellular conditions into the cell, to both the nucleus and cytoplasmic processes to control cell responses. These pathways also engage in various types of signal processing, such as integrating signals over time~\cite{yi2000robust}, converting graded signals to switch-like ones~\cite{ferrell2014ultrasensitivity}, and converting signal strength to signal duration~\cite{behar2008dose}. In eukaryotes, these pathways tend to be highly interconnected, encompassing cross-talk and signal processing between multiple pathways. Genetic mutations in signaling network components are frequently associated with cancer and can result in cells acquiring the ability to divide and grow uncontrollably. A number of important signaling pathways have been identified as frequently genetically altered in cancers. These include the RTK/RAS/MAP-Kinase pathway, PI3K/Akt signaling, WNT signaling as well as many others. It has been reported that 46\% of cancers are associated with alterations in the RTK/RAS/MAP kinase signaling network~\cite{sanchez2018oncogenic}. Because signaling pathways play such a significant role in cancer initiation and advancement, signaling pathways offer an attractive therapeutic target.

Conventional targeted drug discovery, in which a drug is designed to inhibit a specific gene product in a specific signaling pathway, has rarely led to substantial therapeutic effects for complex diseases such as cancer~\cite{aggarwal2007targeting}. These failures arise from poor target specificity, undesirable side effects, and acquired resistance, many of which can be traced to an inadequate understanding of the drug's impact on the overall signaling pathway. Quantitative pathway modeling offers a partial solution to these problems because it can provide a deeper understanding of the impact on the interconnections of the cellular systems being targeted. It can show how a drug affects an entire signaling pathway and cross-talk to other pathways. It can also show how sensitive a pathway's output is to many possible perturbations, which can help identify drug targets as well as synergistic multi-target treatments. 

As an example, several antibacterial drugs have been developed to target the LpxC protein in \textit{E. coli}, which is a natural control point in the metabolism of lipopolysaccharide, an essential cell wall component. However, progress has been hampered by rapid evolution of pathogen resistance~\cite{walsh2014prospects}. Subsequent modeling~\cite{emiola2015complete} showed that feedback in this system maintains constant flux through this control point, which reduces drug effectiveness and offers multiple routes for resistance. This modeling also identified other targets that are likely to be more effective.

This raises the question of whether signaling pathway modeling can assist with identifying effective drugs for treating diseases, such as cancer. An achievement that would facilitate this goal would be identifying sensitive biochemical parameters in signaling pathways, such as molecule abundances and chemical reaction rates. Such information would help determine effective points of attack by therapeutics. If a quantitative model includes a relevant target phenotype, it could be used to identify the most effective points where drugs should act.

In this review we describe work that has been done in recent years in understanding the dynamics and sensitivity of signaling networks that are found in eukaryotic systems. Unfortunately the literature is not extensive and is a topic that has been somewhat neglected by the community. We therefore review the current state of the field together with possible future challenges and areas that need more attention. 

\section{Basic Concepts} \label{sec:1}

Signaling pathways exist at a wide range of complexities. Bacterial signaling is dominated by two-component regulatory systems, in which receptors phosphorylate and dephosphorylate downstream response regulators, and those regulators then transmit the signal on to the necessary targets. However, these types of systems are rare in eukaryotes, being replaced by more complex signaling systems such as multi-step kinase cascades with dual phosphorylation steps. Indeed, a large fraction of signaling networks contain phosphorylation/dephosphorylation cycles. The phosphorylation step is often ATP dependent and is catalyzed by a kinase. The dephosphorylation step is a simple release of free phosphate catalyzed by a phosphatase.

\begin{figure}[htb]
\centering
\includegraphics[scale=0.35]{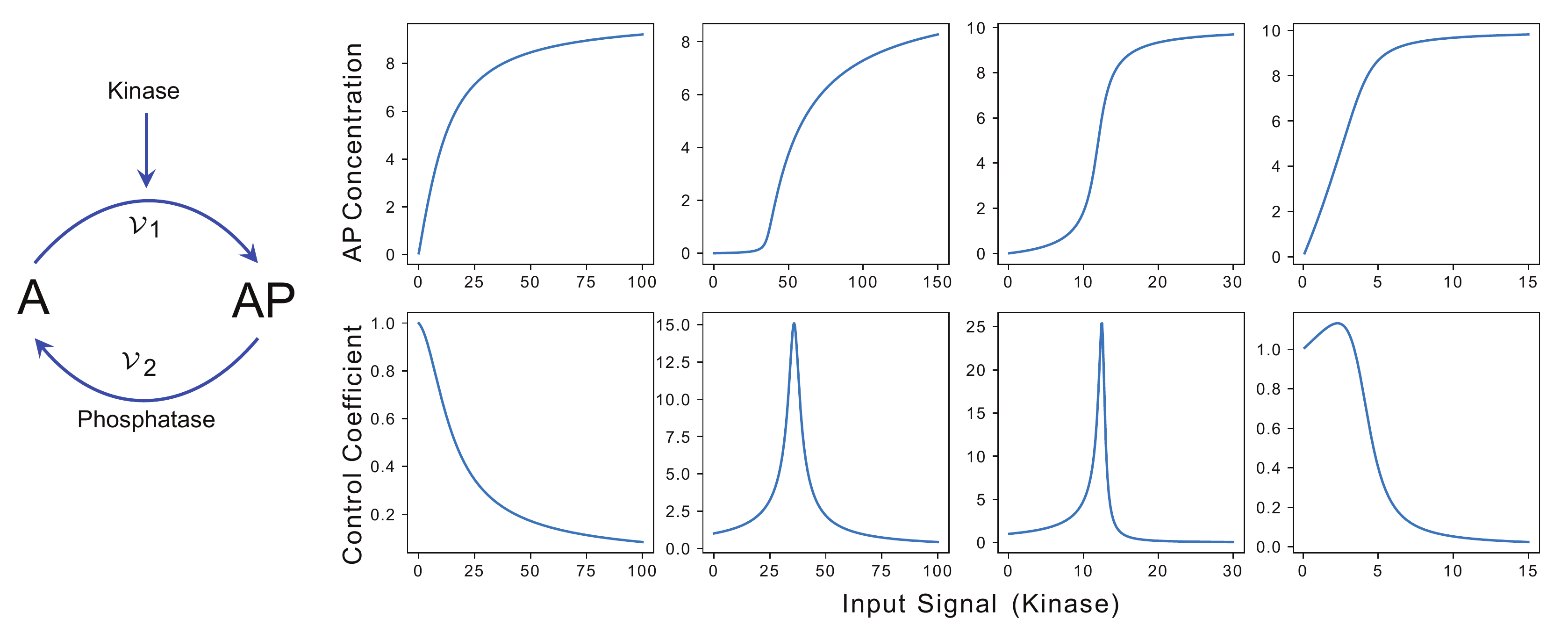}
\caption{Four regimes~\cite{gomez2007operating} of behavior for a simple phosphorylation cycle (left) where $A + AP = T$, $v_1$ is the forward rate catalyzed by a kinase and $v_2$ the reverse rate catalyzed by a phosphatase. The $x$-axis represent changes in kinase activity. {\bf Top Panel:} Hyperbolic (Both Michaelis-Menten constants, K$_m$ are above $A$ and $AP$ concentrations), Threshold Hyperbolic (Kinase $K_m$ above $A$, phosphatase $K_m$ below $AP$), Sigmoid (Both $K_m$s is below $A$ and $AP$ concentrations) and Linear (Kinase $K_m$ below $A$, phosphatase $K_m$ above $AP$). {\bf Bottom panel:} Scaled sensitivity of the species $AP$ with respect to change in the kinase. Code available: See end of article.}
\label{fig:SimpleConservedCycle}
\end{figure}

Figure~\ref{fig:SimpleConservedCycle} shows a typical cycle where the ATP dependence and loss of phosphate has been omitted for clarity. In the remainder of the article we will designate $A$ to represent unphosphorylated protein and $AP$ phosphorylated protein.   An important property of phosphorylation cycles emerges if we assume that protein synthesis and degradation rates are small compared to the phosphorylation and dephosphorylation rates. Under these conditions, the total mass of the cycle is fixed, that is $A + AP = \mbox{constant} = T$. The symbol $T$ will be used to indicate the total mass in a given cycle.

The steady-state properties of such cycles are well known~\cite{stadtman1977superiority}. For example, increasing kinase concentrations shift the steady-state toward $AP$, and vice versa for phosphatase concentrations. $AP$ can respond to kinase concentration changes with hyperbolic, sigmoidal, or linear behavior. Hyperbolic behavior occurs when the $A$ and $AP$ substrate concentrations are below the $K_m$ values for the kinase and phosphatase enzymes, respectively, which can arise from low substrate concentrations, high enzyme concentrations, or fast dissociation of the enzyme-substrate complex. When enzyme saturation values are low, reaction rates are proportional to their substrate concentrations. Hyperbolic behavior also occurs even if the kinase reaction is saturated~\cite{gomez2007operating}. Sigmoidal behavior occurs when both substrate concentrations are above their respective $K_m$ values and enzyme saturation occurs. In the limit of high saturation, both reaction rates are independent of their substrate concentrations, so the system shifts to being entirely $A$ or entirely $AP$, depending on which reaction is faster. Finally, linear behavior arises if the phosphatase reaction is saturated but not the kinase reaction.

Metabolic control analysis~\cite{fell1992metabolic} can help clarify these behaviors. Here, the elasticity of a reaction, $\epsilon^{reaction}_{substrate}$, represents the variation of the reaction rate with respect to its substrate concentration, with the rough interpretation that it is the percent reaction rate increase upon a 1\% substrate concentration increase. The elasticities for the kinase and phosphatase reactions are given by
\begin{align}
\el{1}{A}&=\frac{\partial v_1}{\partial A}\frac{A}{v_1},\\  
\el{2}{AP}&=\frac{\partial v_2}{\partial AP}\frac{AP}{v_2}.
\end{align}
For an irreversible Michaelis-Menten mechanism, elasticities range from 0 for fully saturated enzymes to 1 for unsaturated enzymes. The control coefficient, $C^{substrate}_{enzyme}$, represents the variation of a substrate concentration with respect to an enzyme concentration. For a phosphorylation cycle, the sensitivity of the phosphorylated form $AP$ with respect to the kinase activity has been shown to be \cite{small1990covalent,kholodenko1997quantification,sauro2004quantitative}
\begin{equation}
C^{AP}_{k} = \frac{M_{A}}{M_{AP}\ \el{1}{A} + M_{A}\ \el{2}{AP}},
\label{eqn:CycleSensitivity}
\end{equation}
where $\varepsilon$ are the elasticities, $M$, the mole fraction and $C^{AP}_{k}$, the control coefficient of $AP$ with respect to the kinase concentration. When the concentration of $A$ and $AP$ are above the $K_m$s of the kinase and phosphatase, the elasticities will be small in magnitude. This results in a large sensitivity and corresponds to the case when the cycle is operating in the sigmoid regime. In engineering terminology, the coefficient $C^{AP}_{e_1}$, can be equated to the gain of the cycle between the kinase and output. For example, if both elasticities equal $0.1$, and $90\%$ of the mass is in the form of $A$, then the gain will equal 9, implying that a one percent increase in kinase activity will lead to a $9\%$ increase in the phosphorylated form $AP$. This kind of increased sensitivity is called zero-order ultrasensitivity~\cite{goldbeter1981amplified}.

Cascades of phosphorylation cycles, where the output of one cycle is the kinase for the next cycle, are observed in nature leading to a compounding of the overall gain~\cite{huang1996ultrasensitivity}. The pathway gain increases as a simple product of the separate cycle gains~\cite{brown1997protein,kholodenko1997quantification}. For example, if a cascade is made up of three cycles and each cycle shows a gain of $9$, then the overall gain of the cascade will be $729$. It is clear that substantial amplification can be observed when combining cycles into layers.

Kholodenko et al.~\cite{kholodenko1997quantification} showed that if we ignore sequestration, it is possible to define a unit amplification term $r^i_j$, where for a cascade of $n$ layers, the overall sensitivity is given by
\begin{equation}
R^{p_n}_s = r^{p_1}_s r^{p_2}_{p_1} r^{p_3}_{p_2} \hdots r^{p_n}_{p_{n-1}},
\label{eqn:totalSensitvity}
\end{equation}
where $r^i_j$ is defined the response given in equation~\eqref{eqn:CycleSensitivity} multiplied by the input elasticity from an earlier state stage: $\varepsilon^v_{p_1}$. Once there is sequestration between layers, the simple relationship~\eqref{eqn:totalSensitvity} no longer holds~\cite{bluthgen2006effects}. 

\begin{figure}
    \centering
    \includegraphics[scale=0.5]{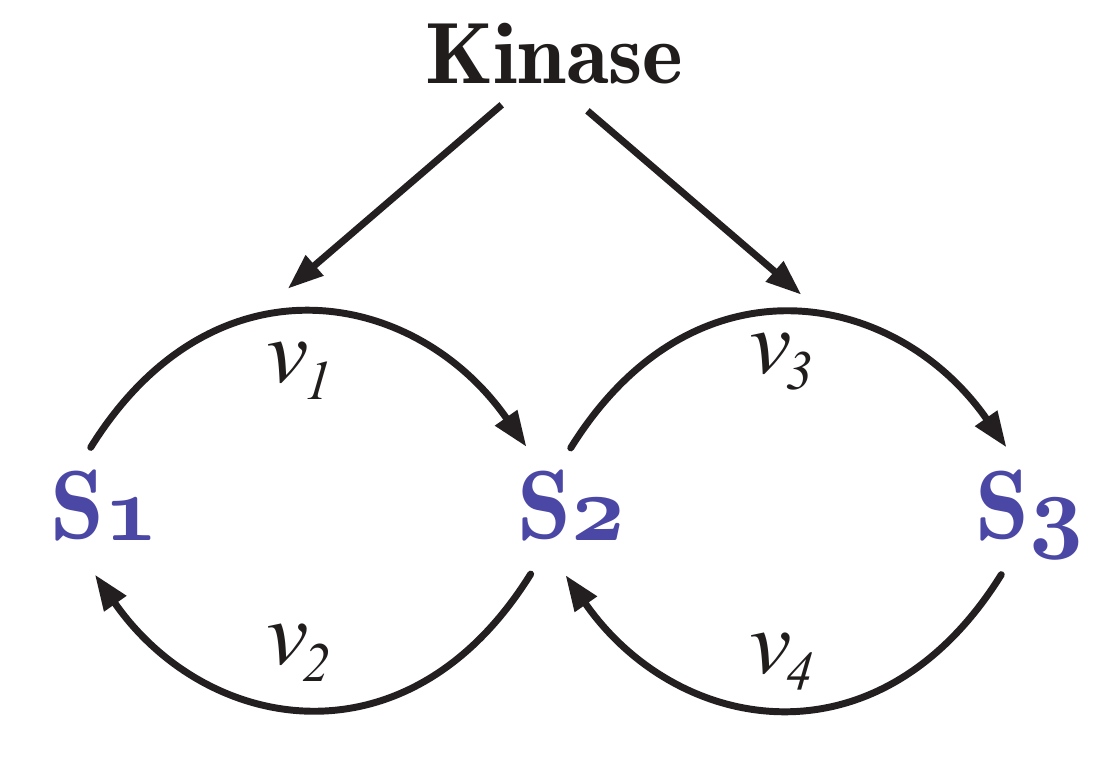}
    \caption{Double phosphorylation cycle.}
    \label{fig:doublecycle}
\end{figure}

In addition to simple cycles, we also find doubly phosphorylated cycles in signaling pathways, particularly in the case of mitogen-activated protein kinase (MAPK) cascades. Figure~\ref{fig:doublecycle} illustrates a double cycle. In doubly-phosphorylated cycles, it is possible to achieve gains greater than one without the concentrations of the proteins going above the $K_m$~\cite{huang1996ultrasensitivity}. However, the gain under these circumstances is limited to the number of phosphorylation sites (and can also be substantially less~\cite{gunawardena2005multisite}). This kind of increased sensitivity is called first-order sensitivity because it arises when there is no saturation of the kinases or phosphatases. It is also possible to combine zero-order together with first-order to generate even higher gains~\cite{markevich2004signaling}.

\begin{figure}
    \centering
    \includegraphics[scale=0.5]{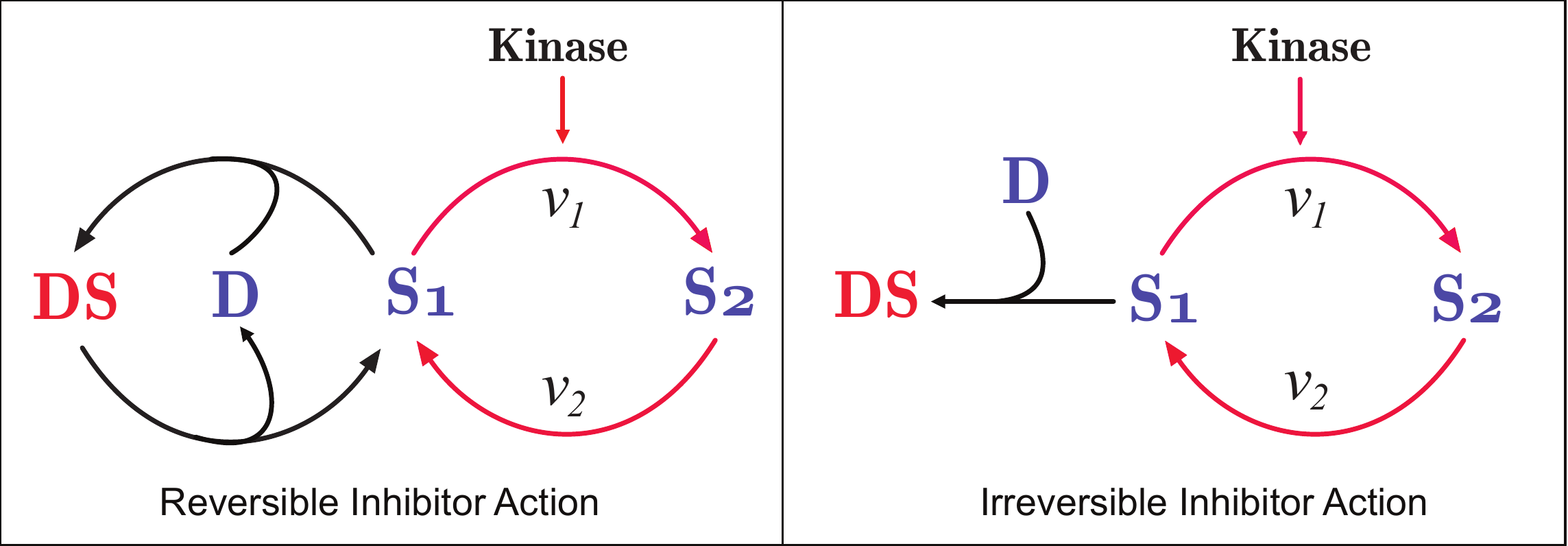}
    \caption{Two possible modes of inhibitor action. $D$ represents the inhibitor. Binding of inhibitor can be irreversible or reversible.}
    \label{fig:DrugActionSingleCycle}
\end{figure}

In terms of drug action, one can imagine at least two scenarios on phosphorylation cycles. A drug could bind irreversibly to either $A$ or $AP$ thus reducing the total mass, $T$, of the cycle; or a drug may bind reversibly to $A$ or $AP$ resulting in competition between the drug and binding of the kinase. These two cases are shown in Figure~\ref{fig:DrugActionSingleCycle}

Investigating the sensitivity of a phosphorylation cycle to changes in $T$ is straight forward with~\cite{sauro1994moiety,Sauro:MCABook}
\begin{equation}
C^{AP}_T = \frac{\varepsilon_1}{M_A \varepsilon_1 + M_{AP} \varepsilon_2}.
\label{eqn:Tsensitivity}
\end{equation}
If the kinase and phosphatase are unsaturated then the elasticities will be roughly one. This means that $C^{AP}_T$ will also be roughly equal to one. That is, when an irreversible inhibitor acts at a particular phosphorylation cycle it has a proportional effect on the output of the cycle. However if the cycle is part of a layer of cycles then this inhibition will be transmitted to downstream cycles and potentially amplified. 

There have not been studies to-date on the effect of reversible inhibition (left panel~\ref{fig:DrugActionSingleCycle}) on signaling cascades.

\section{Feedback and feedforward}

Negative feedback loops are common components in signaling systems. In many cases, negative feedback loops are caused by activated downstream proteins inhibiting the activity of upstream proteins. Examples are seen in the Wnt~\cite{lustig2002negative}, EGFR-Erk~\cite{sturm2010mammalian}, and TGF-$\beta$~\cite{stroschein1999negative,miyazono2000positive} pathways. However, negative feedback can also occur through signal propagation back up the main signaling pathway due to either product inhibition~\cite{cornish2001information} arising from reaction reversibility, or enzyme sequestration~\cite{ventura2008hidden,del2008modular}. Negative feedback always reduces the signaling network gain, which is presumably undesirable in many cases. However, this cost is offset by the numerous control opportunities that negative feedback provides.

\begin{figure}
    \centering
    \includegraphics{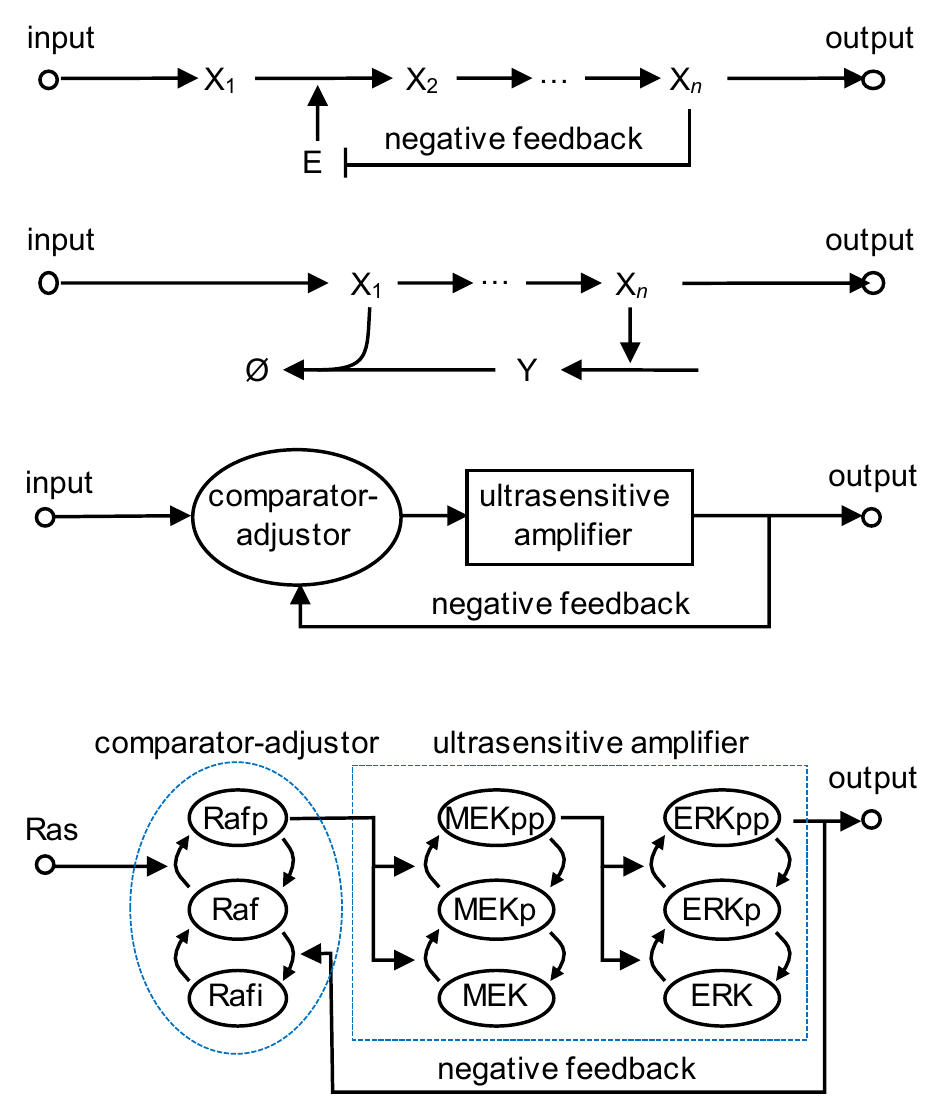}
    \caption{Negative feedback applications. (A) Simple negative feedback to an enzyme to keep concentrations of $X_2$ to $X_n$ and the output nearly constant. (B) Antithetic integral control, which creates perfect adaptation of the output. (C) Feedback for linear signaling from input to output. (D) A specific example of linear signaling using negative feedback, here in the Ras-ERK pathway.}
    \label{fig:feedbackfigure}
\end{figure}

First, negative feedback is widely used in both engineered and biological systems to maintain specific system parameters at near-constant levels, independent of external perturbations and with reduced sensitivity to noise~\cite{bechhoefer2005feedback,sauro2017control,rao2002control}. Biochemical mechanisms for this can be quite simple, such as a downstream protein that inhibits an upstream enzyme (Figure~\ref{fig:feedbackfigure}A). As an example, the glycolysis metabolic pathway maintains constant intracellular ATP concentrations by using negative feedback to control the flux through the phosphofructokinase enzyme~\cite{sauro2017control}, where flux is reduced if ATP concentrations are too high and increased if ATP concentrations are too low. This type of feedback is relatively insensitive to its specific parameters. However, the particular value of that steady-state output does depend on the parameters. Negative feedback takes time to reach a new steady state when the network is perturbed. For example, if the input value is at one constant value and is then stepped up to a higher value, this increase propagates through the system to create an increased output. The negative feedback then takes effect, bringing the output value back down again to adapt the system to the new input value.

More precise adaptation can be achieved through integral control negative feedback, in which the difference between the system's input and output is integrated over time, and the value of that integral is then fed back to reduce the output. This type of control can produce ``perfect adaptation'', meaning that after the input changes, the output adapts exactly back to its previous level. Integral control was first recognized in the \textit{E. coli} chemotaxis pathway~\cite{yi2000robust} and has also been found in the yeast osmoregulation~\cite{muzzey2009systems} system. A particularly simple implementation of integral feedback control is through ``antithetic integral control,'' in which the system generates molecules at both its input and output, each at constant rates, and these molecules bind tightly to each other (Figure~\ref{fig:feedbackfigure}B). The remaining unbound molecules represent the integral of the difference between input and output.

A separate role of negative feedback is to linearize the outputs of signaling systems, making them directly proportional to their inputs. This use has been well-known in electronics for almost a century~\cite{black1934stabilized} but was only recently discovered to operate in biochemical networks~\cite{sauro2007mapk,nevozhay2009negative,andrews2016push,nunns2018signaling}. Achieving linearity in biochemical networks is more challenging than it might appear because all signals asymptotically approach some maximum level, and most simple networks approach this level gradually, resulting in hyperbolic or sigmoidal response curves (see Figure~\ref{fig:SimpleConservedCycle}). Linear response curves are possible from single cycles as well, also shown in Figure~\ref{fig:SimpleConservedCycle}, but these cycles become non-linear when linked in series. This is because when the intermediate enzymes are saturated for linear signaling, it sequesters them from the prior cycle, which makes that one non-linear~\cite{andrews2016push}. Negative feedback can solve this biochemical linearity problem in the same way as in electronic systems. Here, the downstream signal is fed back to a ``comparator-adjustor'' that compares the output to the input and then sends the difference value to an ultrasensitive amplifier to generate the output (Figure~\ref{fig:feedbackfigure}C). The Ras-ERK pathway uses this mechanism (Figure~\ref{fig:feedbackfigure}D). Here, Ras activates Raf, which acts an a comparator-adjustor; Raf is phosphorylated (activated) by an input signal and phosphorylated (inactivated) by a negative feedback signal, with the result that the net concentration of activated Raf represents the difference between system input and output (Figure~\ref{fig:feedbackfigure}D). This value is then amplified through an ultrasensitive kinase cascade~\cite{nunns2018signaling,andrews2018signaling}.

Systems with negative feedback loops generally have a tendency to oscillate due to the feedback always acting in opposition to the output. Actual oscillations depend on the feedback strength, system gain, and damping forces within the system~\cite{bechhoefer2005feedback}. Damping forces in physical oscillators typically arise from energy dissipating influences, such as mechanical friction or electrical resistance. In biochemical oscillators, they often arise instead from the stochastic nature of the biochemical reaction, which removes phase coherence and hence damps oscillations; here, energy dissipation is required to \textit{maintain} oscillations~\cite{cao2015free}. The configuration of some cell signaling networks can also dampen oscillations~\cite{hoffmann2002ikappab}. Biochemical oscillations in cells help control cell cycles and circadian rhythms, but there are cases where their purpose is less clear , including in glycolysis~\cite{boiteux1975control}, the EGFR-ERK signaling system~\cite{shankaran2009rapid}, and the NF-$\kappa$B system~\cite{hoffmann2002ikappab}. Various roles for these oscillations have been proposed~\cite{boiteux1975control,shankaran2010oscillatory}, but they could also be, at least partially, simply a side effect of negative feedback that has another important regulatory role.

Positive feedback loops also connect elements from downstream parts of a network to upstream ones, but act to increase the downstream signal. Their steady-state impact is to increase the gain of a signaling pathway, meaning that they compress the dose-response curve. For example, the MAPK cascade in Xenopus oocytes includes a positive feedback loop that increases the Hill coefficient of the dose-response curve from 5 to 35, essentially creating an all-or-none response~\cite{ferrell1998biochemical}. Even stronger positive feedback converts the dose-response curve from a steep sigmoid to an S-shaped curve containing a region of bistability~\cite{ferrell2002self,xiong2003positive}. Here, there is a specific input range over which the signaling system output is stable with either high or low outputs. Moving outside of the stable range can flip the system between the two alternative states. Although bistability require positive feedback, it doesn't have to be effected through an external circuit element. For example, enzyme saturation and competitive inhibition are sufficient in cycles that utilize dual phosphorylation mechanisms~\cite{markevich2004signaling}. Bistability is a necessary condition for relaxation oscillators. Additionally, if the system's bistable range covers all possible input parameters, then the output cannot be flipped between states, but is irreversibly locked into one state or the other. This behavior is useful for irreversible cell state changes, such as for one-way progression through the cell cycle, and irreversible cell fate decisions during differentiation. The lambda-phage lysis/lysogeny decision~\cite{arkin1998stochastic} is a particularly well-studied example of positive feedback being used to lock in a particular fate decision.

Feedforward elements in cell signaling transmit the signal downstream parallel to the main signaling pathway followed by signal recombination. Positive feedforward loops, in which the feedforward signal has the same sign as the main signal, are called coherent, while negative feedforward loops that have the opposite effect are called incoherent~\cite{shen2002network}. Feedforwards have been studied much less than feedbacks but share some of the same control behaviors. In particular, incoherent feedforward loops can perform perfect adaptation~\cite{ma2009defining,sontag2009remarks} much like integral feedback. This was observed experimentally in a Ras signaling pathway used for \textit{Dictyostelium discoideum} chemotaxis, where the signal was sent to two proteins simultaneously, with one activating a downstream Ras protein and the other inhibiting it~\cite{takeda2012incoherent}.

\begin{figure}
    \centering
    \includegraphics{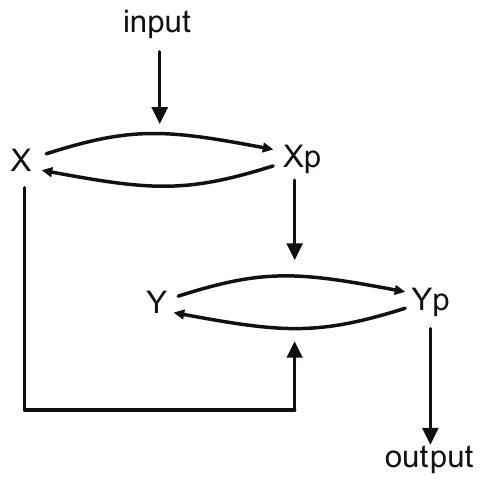}
    \caption{Diagram of a push-pull mechanism.}
    \label{fig:PushPull}
\end{figure}

A particularly interesting type of feedforward loop is a push-pull mechanism~\cite{chock1977superiority,andrews2016push}. In a simple version (Figure~\ref{fig:PushPull}), there are two proteins, here called X and Y, each of which cycle between inactive and active states which are shown in the figure as unphosphorylated or phosphorylated states. The active copies of X phosphorylate Y, pushing it toward greater activity, as is typical in kinase cascades. Meanwhile, the inactive X dephosphorylates Y, pulling it back toward inactivity. Push-pull mechanisms are often described as paradoxical signaling components or incoherent feedforward loops~\cite{hart2013utility,rangamani2016paradoxical} because the same X protein acts to both activate and inactivate Y, but these descriptions are misleading because the behaviors arise from different states of the X protein; indeed, it would be more appropriate to call it a coherent feedforward loop due to the double sign reversal in the feedforward loop. The symmetry of the push-pull mechanism gives it several unique properties. It is the only type of feedback or feedforward that can transmit signals linearly when the relevant enzymes are unsaturated~\cite{andrews2016push}. It also confers substantial robustness to variability in the system's components~\cite{shinar2007input} and is able to provide ratiometric sensing rather than absolute sensing, meaning that the activity of Y depends on the \textit{fraction} rather than the amount of X that is active~\cite{bush2016yeast}.

Many push-pull mechanisms have been identified, including bacterial two-component signaling~\cite{batchelor2003robustness,shinar2007input}, the yeast pheromone response receptors~\cite{bush2016yeast}, the yeast MAP kinase pathway~\cite{andrews2016push}, and the heat shock response in Xenopus oocytes~\cite{conde2009modulation}.

\section{Effects of Sequestration}

Ultrasensitivity in signaling pathways takes the form of sharp sigmoidal responses to small changes in an input signal (Figure~\ref{fig:SimpleConservedCycle}) and allows for threshold-based switch-like behavior. With respect to phosphorylation/dephosphorylation, or any other covalent modification cycle, the 'speed' at which this switch-like transition happens, as the kinase level is increased, can be quantified with a response coefficient defined, as in~\cite{goldbeter1981amplified}, as
\begin{equation}
R_{c} = \frac{K/P\text{ at }90\% \, AP}{K/P\text{ at }10\% \, AP},
\label{eqn:eqn3}
\end{equation}
where \(K/P\) is the ratio of the kinase and phosphatase enzymes and \(AP\) is phosphorylated form of the protein.  An alternative approach to the response coefficient in equation~\eqref{eqn:eqn3} is one that makes use of metabolic control analysis as discussed above~\eqref{fig:SimpleConservedCycle}~\cite{small1990covalent}. Note that in equation~\eqref{eqn:CycleSensitivity} there are no terms for enzyme-substrate complexes. The derivation of that equation assumes that the concentrations of the modified and unmodified substrates are much greater than the enzymes that target them. The amount of bound substrate is thus negligible, allowing enzyme-substrate complexes to be ignored.

However, enzymes and their targets often have comparable cellular concentrations~\cite{albe1990cellular,bluthgen2006effects} as in the MAPK cascade. Increasing enzyme concentrations to levels on par with their target substrates may lower the concentration of free substrates to levels below their $K_m$ values~\cite{bluthgen2006effects}, a condition for ultrasensitivity. Substrate sequestration could also render them unavailable to other enzymes in the pathway. It is therefore necessary to consider the role of substrates sequestration in the overall behavior of signalling pathways.

Inclusion of enzyme-substrate complexes in mechanistic computational models~\cite{goldbeter1981amplified} has demonstrated that increasing enzyme concentrations, and subsequently increasing the levels of enzyme-substrate complexes, results in a significant reduction in sensitivity as the ratio of enzyme to total substrate grows large. Another analysis approach is to extend the existing MCA-based sensitivity~\eqref{eqn:CycleSensitivity} to accommodate the additional enzyme-substrate complex components~\cite{bluthgen2006effects}. The derived result is
\begin{equation}
C^{AP}_{k} = \frac{M_{A}}{M_{AP}\ \el{1}{A} + M_{A}\ \el{2}{AP}  + M_{K*A}\el{1}{A}\el{2}{AP} + M_{P*AP}},
\end{equation}
where the new terms \(M_{K*A}\) and \(M_{P*AP}\) are the mole fractions of bound kinase and phosphatase respectively.
The end result is two-fold. If we assume the elasticities remain constant, the shift in mass of \(A\) to the denominator, and the shift in mass of \(AP\) to a term in the denominator lacking an elasticity multiplier, will lower sensitivity. It is also clear that if we have sufficient concentrations of enzymes such that we cannot disregard the enzyme-substrate complexes, sequestration will lower the concentrations of free substrate and result in higher elasticities. This in turn also lowers sensitivities.

\begin{figure}
    \centering
    \includegraphics[scale=0.8]{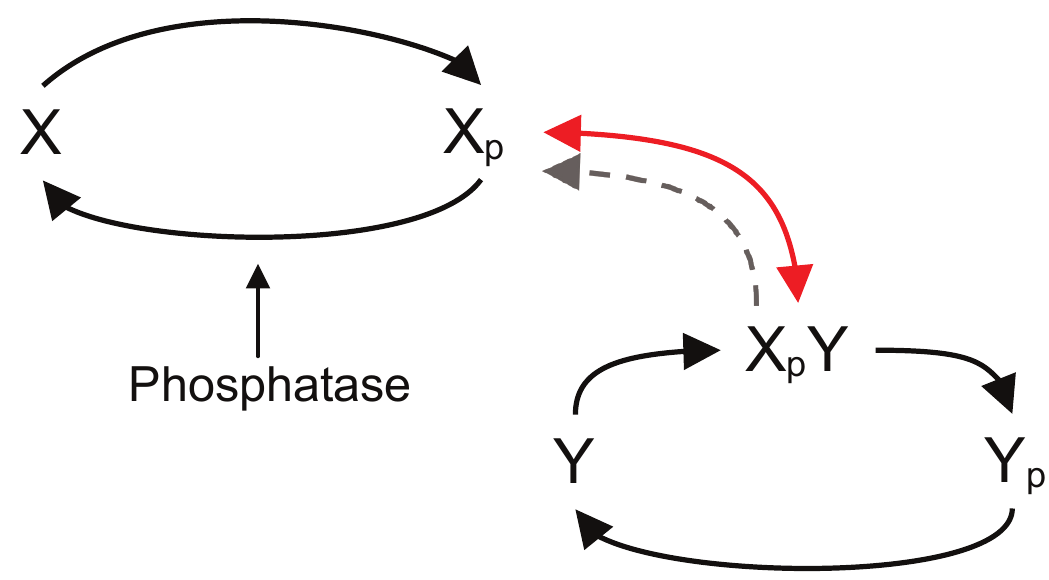}
    \caption{Retroactivity in sequential phosphorylation cycles. The phosphorylated species in the upstream cycle (\(X_p\)) acts as a kinase in the downstream cycle (red arrow). This interaction induces indirect upstream effects (dashed gray arrow) on the upstream cycle by sequestration of the substrate for the upstream phosphatase.}
    \label{fig:feedbackfigureRetroactivity}
\end{figure}

Higher enzyme levels and substrate sequestration carry clear ramifications for signaling pathway dynamics. The four operating regimes shown in Figure~\ref{fig:SimpleConservedCycle} consider the case in which the substrate concentrations are much greater than that of the enzymes. By allowing the enzyme concentrations to increase to levels on par with, or greater than, the substrate, the number of operating regimes greatly expands~\cite{straube2017operating}. The four operating regimes in Figure~\ref{fig:SimpleConservedCycle} have also been analyzed in the context of downstream loads on the system~\cite{parundekar2020retroactivity}. Substrates are often enzymes in their own right and can act on targets that alter upstream behavior of the system, a process termed retroactivity (Figure~\ref{fig:feedbackfigureRetroactivity})~\cite{del2008modular}. These retroactivity effects induce five distinct transitions between the aforementioned operating regimes depending on the initial regime and the substrate (phosphorylated or unphosphorylated) taking part in downstream interactions.

Retroactivity brought about via downstream substrate sequestration is a key behavioral consideration in covalent modification-based signaling pathways~\cite{wynn2011kinase,sepulchre2012retroactive,ossareh2011long} like the MAPK cascade. This was demonstrated in~\cite{ventura2008hidden} with a computational model of a three-level cascade of phosphorylation cycles that incorporated the effects of enzyme-substrate complexes. This showed that that signal responses for the two intermediate cycles had lower maximum steady-state values of the active substrate as compared to the final cycle that had no downstream targets. This implies there is sequestration of the intermediate active substrates from their role as enzymes by downstream targets. There was also a reduction in sensitivity when compared to a model that doesn't include the effects of sequestration~\cite{goldbeter1981amplified}. Retroactivity has also been demonstrated experimentally, for example, in~\cite{ventura2010signaling} with a uridylylation/deuridylylation cycle derived from \textit{E. coli}. It was shown that in the presence of a downstream binding target (NRII) for the substrate (PII) the sensitivity to an increase in a regulator (glutamine) was diminished in comparison to the case lacking a downstream target. It has also been shown that substrates of phosphorylated MAPK, the last phosphorylation cycle in the MAPK cascade, inhibited subsequent MAPK dephosphorylation~\cite{kim2011substrate}.

Substrate sequestration and its potential roles in system dynamics have been explored, computationally, in the context of various mechanistic models. For example, phosphatase sequestration in the presence of an opposing fast, low affinity kinase can cause signal desensitization, while sequestration in the context of a cycle with a doubly phosphorylated substrate can result in a signal with a sign-sensitive delay \cite{bluthgen2006effects}. These two scenarios potentially obviate the need for negative feedback and coherent feedforward loops respectively. Substrate sequestration combined with multiple substrate phosphorylation sites can also be used to design modules with optimal threshold and ultrasensitive responses as detailed in~\cite{liu2010combination}. Under the right conditions ultrasensitivity in a phosphorylation cycle may be achieved even when the enzymes are unsaturated. As demonstrated in \cite{martins2013ultrasensitivity}, when the kinase and phosphatase require docking to a separate site before catalysis the docking site can serve as a means for the enzyme to sequester the substrate from the opposing enzyme. For example a kinase binding to a fully phosphorylated substrate could block the opposing phosphatase. In this way, sequestering of the  substrate would enable ultrasensitivity rather than destroying it.

Bistability in double phosphorylation cycles (Figure~\ref{fig:doublecycle}) emerges from the saturation of at least one of the enzymes by their fully phosphorylated/dephosphorylated substrate~\cite{markevich2004signaling}. However, accounting for substrate sequestration by incorporating the complexes into the equations produced a reduced parameter domain (e.g., enzyme $K_m$ values) on which bistability is possible. In addition, when one enzyme is left unsaturated the bistability domain can shrink considerably. Sequestration also has a deleterious effect on the stability of steady states in cascades of double phosphorylation cycles~\cite{zumsande2010bifurcations} since stability is negatively correlated to high levels of enzyme bound substrate in the last cycle of the cascade. The bistability conditions laid out in~\cite{markevich2004signaling} were subsequently combined with negative feedback loops \cite{chickarmane2007oscillatory} to construct relaxation and ring oscillators, while work in~\cite{jolley2012design} further showed that oscillations are achievable with just a double phosphorylation cycle (Figure~\ref{fig:doublecycle}) within certain parameter ranges and, in general, an excess of substrate over enzyme. Nevertheless, increasing enzyme concentrations and the subsequent substrate sequestration are know to eliminate oscillatory behavior, as was demonstrated in \cite{bluthgen2006effects} using an oscillatory model of the MAPK cascade along with a negative feedback loop.

The majority of the work on substrate sequestration has been computational, and thus hypothetical. In this section we have discussed some of the more important aspects of sequestration as well as a few of its potential roles. Additional computational studies with a broader perspective on the topic can be found in~\cite{suwanmajo2015mixed,feng2016enzyme,RN414,sarma2012different}. A limited number of cases for zero-order sensitivity have been established experimentally, for which sequestration could potentially play a role, including the phosphorylation cycles of glycogen phosphorylase \cite{meinke1986zero} and isocitrate dehydrogenase \cite{laporte1983phosphorylation}. Of course the general concept of sequestration extends much more broadly than substrate sequestration by enzymes in covalent modification cycle. Examples include the hypothesis that the bistability in the Raf-Mek-Erk signal cascade is governed by the sequestration of Mek by Erk resulting in a positive feedback loop \cite{legewie2007competing} and the sequestration of enzymes via pathway competition as demonstrated with a model system of the Jak kinase \cite{dondi2001down}.

\section{Discussion}

In this review, we described how the intrinsic properties of enzymes impact the sensitivity and dynamics of signaling networks. This knowledge is essential in predicting how drugs and genetic perturbations will impact those networks. Knowledge of how these networks operate as an integrated system, however, is very incomplete. For example, we have a reasonably good understanding of the enzymology of the three component Raf-Mek-Erk cascade (Figure~\ref{fig:feedbackfigure}), but the upstream and downstream components are far less understood. Because substrate sequestration can have a profound impact on the function of signaling networks, this lack of contextual understanding of different components of signaling networks limits our ability to build predictive models of their dynamic behavior. 
There is also a lack of information on the relevant concentrations and distributions of signaling pathway enzymes and substrates in cells. Traditional methods for parameterization of signaling models usually generate “lumped” parameters in which the concentrations of both enzymes and substrates conflated with rate constants~\cite{spencer2010systemic} . This can be useful for modeling the behavior of specific cell system, but such models generally cannot predict the impact of changing enzyme and protein concentrations outside of a limited range. This limits their usefulness for predicting the impact of drugs or genetic changes, such as copy number variations. Several recent studies, however, have quantified pathway protein abundance as explicit model parameters and have shown that this improves the ability of models to predict the impact of both drugs and altered protein abundance~\cite{shi2016conservation}. As the technologies needed to quantify the often-low levels of signaling proteins improve, inclusion of their abundance values in models should improve their predictive power.

A further difficulty in building predictive models of signaling networks is their complexity and interconnectedness. The above examples of motifs and network topologies represent only a small part of signaling networks in cells. Actual signaling networks comprise a complex set of recursive, interconnected pathways that link intercellular information processing with changing extracellular conditions. As outside conditions change, these networks adapt dynamically such that the relative level of their constituents is always in flux. Dealing with this continual rewiring is a challenge that will require new approaches in mathematically representing signaling pathways.

One approach is to represent subsections of these networks as “modules” that explicitly include their dynamic response potential~\cite{joslin2010structure}. Signaling networks are widely thought to be modular due to both the need for evolutionary flexibility in reconfiguring networks in response to selective pressure~\cite{clune2013evolutionary}  and as a mechanism to reduce the impact of biological noise~\cite{atay2014modularity}. An inherent aspect of these modules is that their interaction with other parts of the network should be restricted to just a few “interface” proteins that insulate the modules from other parts of the networks. This insulation can be achieved through negative feedback or low substrate sequestration~\cite{del2008modular}. Thus, understanding the dynamics between different components of a signaling pathways is crucial to understanding their overall regulatory architecture. 

Learning to identify and build models of the constituent modules that comprise signaling networks will provide a powerful new approach for building scalable models of signal transduction. However, understanding the impact on drugs on these networks will always require a fine level understanding of how they impact specific biochemical reactions. Thus, an appreciation of these networks in terms of their underlying biochemistry is crucial in building realistic, and ultimately predictive models of their function in both health and disease.

\begin{acknowledgements}
This work was supported by National Cancer Institute under grant number U01CA242992. The content is solely the responsibility of the authors and does not necessarily represent the official views of the National Institutes of Health, or the University of Washington.
\end{acknowledgements}

\section*{Code Availability}

Code for generating Figure~\ref{fig:SimpleConservedCycle} is available at:~ \url{https://github.com/sys-bio/CodeForPublishedPapers/tree/main/CurrentPathobiologyReports2021}
\section*{Conflict of interest}

The authors declare that they have no conflict of interest.

\bibliographystyle{spmpsci}   
\bibliography{bib.bib}  

\begin{thebibliography}{10}
\providecommand{\url}[1]{{#1}}
\providecommand{\urlprefix}{URL }
\expandafter\ifx\csname urlstyle\endcsname\relax
  \providecommand{\doi}[1]{DOI~\discretionary{}{}{}#1}\else
  \providecommand{\doi}{DOI~\discretionary{}{}{}\begingroup
  \urlstyle{rm}\Url}\fi

\bibitem{aggarwal2007targeting}
Aggarwal, B.B., Sethi, G., Baladandayuthapani, V., Krishnan, S., Shishodia, S.:
  Targeting cell signaling pathways for drug discovery: an old lock needs a new
  key.
\newblock Journal of cellular biochemistry \textbf{102}(3), 580--592 (2007)

\bibitem{albe1990cellular}
Albe, K.R., Butler, M.H., Wright, B.E.: Cellular concentrations of enzymes and
  their substrates.
\newblock Journal of theoretical biology \textbf{143}(2), 163--195 (1990)

\bibitem{andrews2018signaling}
Andrews, S.S., Brent, R., Bal{\'a}zsi, G.: Signaling systems: Transferring
  information without distortion.
\newblock Elife \textbf{7}, e41894 (2018)

\bibitem{andrews2016push}
Andrews, S.S., Peria, W.J., Richard, C.Y., Colman-Lerner, A., Brent, R.:
  Push-pull and feedback mechanisms can align signaling system outputs with
  inputs.
\newblock Cell systems \textbf{3}(5), 444--455 (2016)

\bibitem{arkin1998stochastic}
Arkin, A., Ross, J., McAdams, H.H.: Stochastic kinetic analysis of
  developmental pathway bifurcation in phage $\lambda$-infected escherichia
  coli cells.
\newblock Genetics \textbf{149}(4), 1633--1648 (1998)

\bibitem{atay2014modularity}
Atay, O., Skotheim, J.M.: Modularity and predictability in cell signaling and
  decision making.
\newblock Molecular biology of the cell \textbf{25}(22), 3445--3450 (2014)

\bibitem{batchelor2003robustness}
Batchelor, E., Goulian, M.: Robustness and the cycle of phosphorylation and
  dephosphorylation in a two-component regulatory system.
\newblock Proceedings of the National Academy of Sciences \textbf{100}(2),
  691--696 (2003)

\bibitem{bechhoefer2005feedback}
Bechhoefer, J.: Feedback for physicists: A tutorial essay on control.
\newblock Reviews of modern physics \textbf{77}(3), 783 (2005)

\bibitem{behar2008dose}
Behar, M., Hao, N., Dohlman, H.G., Elston, T.C.: Dose-to-duration encoding and
  signaling beyond saturation in intracellular signaling networks.
\newblock PLoS Comput Biol \textbf{4}(10), e1000197 (2008)

\bibitem{black1934stabilized}
Black, H.S.: Stabilized feedback amplifiers.
\newblock Bell system technical journal \textbf{13}(1), 1--18 (1934)

\bibitem{bluthgen2006effects}
Bl{\"u}thgen, N., Bruggeman, F.J., Legewie, S., Herzel, H., Westerhoff, H.V.,
  Kholodenko, B.N.: Effects of sequestration on signal transduction cascades.
\newblock The FEBS journal \textbf{273}(5), 895--906 (2006)

\bibitem{boiteux1975control}
BoITEUX, A., GOLDBETER, A., HESS, B.: Control of oscillating glycolysis of
  yeast by stochastic, periodic, and steady source of substrate: A model and
  experimental study.
\newblock Proc. Nat. Acad. Sci. USA \textbf{72}(10), 3829--3833 (1975)

\bibitem{brown1997protein}
Brown, G.C., Hoek, J.B., Kholodenko, B.N.: Why do protein kinase cascades have
  more than one level?
\newblock Trends in biochemical sciences \textbf{22}(8), 288--288 (1997)

\bibitem{bush2016yeast}
Bush, A., Vasen, G., Constantinou, A., Dunayevich, P., Patop, I.L., Blaustein,
  M., Colman-Lerner, A.: Yeast gpcr signaling reflects the fraction of occupied
  receptors, not the number.
\newblock Molecular systems biology \textbf{12}(12), 898 (2016)

\bibitem{cao2015free}
Cao, Y., Wang, H., Ouyang, Q., Tu, Y.: The free-energy cost of accurate
  biochemical oscillations.
\newblock Nature physics \textbf{11}(9), 772--778 (2015)

\bibitem{chickarmane2007oscillatory}
Chickarmane, V., Kholodenko, B.N., Sauro, H.M.: Oscillatory dynamics arising
  from competitive inhibition and multisite phosphorylation.
\newblock Journal of theoretical biology \textbf{244}(1), 68--76 (2007)

\bibitem{chock1977superiority}
Chock, P., Stadtman, E.: Superiority of interconvertible enzyme cascades in
  metabolic regulation: analysis of multicyclic systems.
\newblock Proceedings of the National Academy of Sciences \textbf{74}(7),
  2766--2770 (1977)

\bibitem{clune2013evolutionary}
Clune, J., Mouret, J.B., Lipson, H.: The evolutionary origins of modularity.
\newblock Proceedings of the Royal Society b: Biological sciences
  \textbf{280}(1755), 20122863 (2013)

\bibitem{conde2009modulation}
Conde, R., Belak, Z.R., Nair, M., O’Carroll, R.F., Ovsenek, N.: Modulation of
  hsf1 activity by novobiocin and geldanamycin.
\newblock Biochemistry and Cell Biology \textbf{87}(6), 845--851 (2009)

\bibitem{cornish2001information}
Cornish-Bowden, A., C{\'a}rdenas, M.L.: Information transfer in metabolic
  pathways: effects of irreversible steps in computer models.
\newblock European Journal of Biochemistry \textbf{268}(24), 6616--6624 (2001)

\bibitem{del2008modular}
Del~Vecchio, D., Ninfa, A.J., Sontag, E.D.: Modular cell biology: retroactivity
  and insulation.
\newblock Molecular systems biology \textbf{4}(1), 161 (2008)

\bibitem{dondi2001down}
Dondi, E., Pattyn, E., Lutfalla, G., Van~Ostade, X., Uz{\'e}, G., Pellegrini,
  S., Tavernier, J.: Down-modulation of type 1 interferon responses by receptor
  cross-competition for a shared jak kinase.
\newblock Journal of Biological Chemistry \textbf{276}(50), 47004--47012 (2001)

\bibitem{emiola2015complete}
Emiola, A., George, J., Andrews, S.S.: A complete pathway model for lipid a
  biosynthesis in escherichia coli.
\newblock PloS one \textbf{10}(4), e0121216 (2015)

\bibitem{fell1992metabolic}
Fell, D.A.: Metabolic control analysis: a survey of its theoretical and
  experimental development.
\newblock Biochemical Journal \textbf{286}(2), 313--330 (1992)

\bibitem{feng2016enzyme}
Feng, S., Ollivier, J.F., Soyer, O.S.: Enzyme sequestration as a tuning point
  in controlling response dynamics of signalling networks.
\newblock PLoS computational biology \textbf{12}(5), e1004918 (2016)

\bibitem{RN414}
Feng, S., Soyer, O.S.: In silico evolution of signaling networks using
  rule-based models: Bistable response dynamics.
\newblock Methods Mol Biol \textbf{1945}, 315--339 (2019)

\bibitem{ferrell1998biochemical}
Ferrell, J.E., Machleder, E.M.: The biochemical basis of an all-or-none cell
  fate switch in xenopus oocytes.
\newblock Science \textbf{280}(5365), 895--898 (1998)

\bibitem{ferrell2002self}
Ferrell~Jr, J.E.: Self-perpetuating states in signal transduction: positive
  feedback, double-negative feedback and bistability.
\newblock Current opinion in cell biology \textbf{14}(2), 140--148 (2002)

\bibitem{ferrell2014ultrasensitivity}
Ferrell~Jr, J.E., Ha, S.H.: Ultrasensitivity part iii: cascades, bistable
  switches, and oscillators.
\newblock Trends in biochemical sciences \textbf{39}(12), 612--618 (2014)

\bibitem{goldbeter1981amplified}
Goldbeter, A., Koshland, D.E.: An amplified sensitivity arising from covalent
  modification in biological systems.
\newblock Proceedings of the National Academy of Sciences \textbf{78}(11),
  6840--6844 (1981)

\bibitem{gomez2007operating}
Gomez-Uribe, C., Verghese, G.C., Mirny, L.A.: Operating regimes of signaling
  cycles: statics, dynamics, and noise filtering.
\newblock PLoS Comput Biol \textbf{3}(12), e246 (2007)

\bibitem{gunawardena2005multisite}
Gunawardena, J.: Multisite protein phosphorylation makes a good threshold but
  can be a poor switch.
\newblock Proceedings of the National Academy of Sciences \textbf{102}(41),
  14617--14622 (2005)

\bibitem{hart2013utility}
Hart, Y., Alon, U.: The utility of paradoxical components in biological
  circuits.
\newblock Molecular cell \textbf{49}(2), 213--221 (2013)

\bibitem{hoffmann2002ikappab}
Hoffmann, A., Levchenko, A., Scott, M.L., Baltimore, D.: The
  i$\kappa$b-nf-$\kappa$b signaling module: temporal control and selective gene
  activation.
\newblock Science \textbf{298}(5596), 1241--1245 (2002)

\bibitem{huang1996ultrasensitivity}
Huang, C.Y., Ferrell, J.E.: Ultrasensitivity in the mitogen-activated protein
  kinase cascade.
\newblock Proceedings of the National Academy of Sciences \textbf{93}(19),
  10078--10083 (1996)

\bibitem{jolley2012design}
Jolley, C.C., Ode, K.L., Ueda, H.R.: A design principle for a posttranslational
  biochemical oscillator.
\newblock Cell reports \textbf{2}(4), 938--950 (2012)

\bibitem{joslin2010structure}
Joslin, E.J., Shankaran, H., Opresko, L.K., Bollinger, N., Lauffenburger, D.A.,
  Wiley, H.S.: Structure of the egf receptor transactivation circuit integrates
  multiple signals with cell context.
\newblock Molecular BioSystems \textbf{6}(7), 1293--1306 (2010)

\bibitem{kholodenko1997quantification}
Kholodenko, B.N., Hoek, J.B., Westerhoff, H.V., Brown, G.C.: Quantification of
  information transfer via cellular signal transduction pathways.
\newblock FEBS letters \textbf{414}(2), 430--434 (1997)

\bibitem{kim2011substrate}
Kim, Y., Paroush, Z., Nairz, K., Hafen, E., Jim{\'e}nez, G., Shvartsman, S.Y.:
  Substrate-dependent control of mapk phosphorylation in vivo.
\newblock Molecular systems biology \textbf{7}(1), 467 (2011)

\bibitem{laporte1983phosphorylation}
LaPorte, D.C., Koshland, D.E.: Phosphorylation of isocitrate dehydrogenase as a
  demonstration of enhanced sensitivity in covalent regulation.
\newblock Nature \textbf{305}(5932), 286--290 (1983)

\bibitem{legewie2007competing}
Legewie, S., Schoeberl, B., Bl{\"u}thgen, N., Herzel, H.: Competing docking
  interactions can bring about bistability in the mapk cascade.
\newblock Biophysical journal \textbf{93}(7), 2279--2288 (2007)

\bibitem{liu2010combination}
Liu, X., Bardwell, L., Nie, Q.: A combination of multisite phosphorylation and
  substrate sequestration produces switchlike responses.
\newblock Biophysical journal \textbf{98}(8), 1396--1407 (2010)

\bibitem{lustig2002negative}
Lustig, B., Jerchow, B., Sachs, M., Weiler, S., Pietsch, T., Karsten, U.,
  van~de Wetering, M., Clevers, H., Schlag, P.M., Birchmeier, W., et~al.:
  Negative feedback loop of wnt signaling through upregulation of
  conductin/axin2 in colorectal and liver tumors.
\newblock Molecular and cellular biology \textbf{22}(4), 1184--1193 (2002)

\bibitem{ma2009defining}
Ma, W., Trusina, A., El-Samad, H., Lim, W.A., Tang, C.: Defining network
  topologies that can achieve biochemical adaptation.
\newblock Cell \textbf{138}(4), 760--773 (2009)

\bibitem{markevich2004signaling}
Markevich, N.I., Hoek, J.B., Kholodenko, B.N.: Signaling switches and
  bistability arising from multisite phosphorylation in protein kinase
  cascades.
\newblock The Journal of cell biology \textbf{164}(3), 353--359 (2004)

\bibitem{martins2013ultrasensitivity}
Martins, B.M., Swain, P.S.: Ultrasensitivity in
  phosphorylation-dephosphorylation cycles with little substrate.
\newblock PLoS Comput Biol \textbf{9}(8), e1003175 (2013)

\bibitem{meinke1986zero}
Meinke, M.H., Bishop, J.S., Edstrom, R.D.: Zero-order ultrasensitivity in the
  regulation of glycogen phosphorylase.
\newblock Proceedings of the National Academy of Sciences \textbf{83}(9),
  2865--2868 (1986)

\bibitem{miyazono2000positive}
Miyazono, K.: Positive and negative regulation of tgf-beta signaling.
\newblock Journal of cell science \textbf{113}(7), 1101--1109 (2000)

\bibitem{muzzey2009systems}
Muzzey, D., G{\'o}mez-Uribe, C.A., Mettetal, J.T., van Oudenaarden, A.: A
  systems-level analysis of perfect adaptation in yeast osmoregulation.
\newblock Cell \textbf{138}(1), 160--171 (2009)

\bibitem{nevozhay2009negative}
Nevozhay, D., Adams, R.M., Murphy, K.F., Josi{\'c}, K., Bal{\'a}zsi, G.:
  Negative autoregulation linearizes the dose--response and suppresses the
  heterogeneity of gene expression.
\newblock Proceedings of the National Academy of Sciences \textbf{106}(13),
  5123--5128 (2009)

\bibitem{nunns2018signaling}
Nunns, H., Goentoro, L.: Signaling pathways as linear transmitters.
\newblock Elife \textbf{7}, e33617 (2018)

\bibitem{ossareh2011long}
Ossareh, H.R., Ventura, A.C., Merajver, S.D., Del~Vecchio, D.: Long signaling
  cascades tend to attenuate retroactivity.
\newblock Biophysical journal \textbf{100}(7), 1617--1626 (2011)

\bibitem{parundekar2020retroactivity}
Parundekar, A., Viswanathan, G.A.: Retroactivity induced operating regime
  transition in a phosphorylation-dephosphorylation cycle.
\newblock bioRxiv  (2020)

\bibitem{rangamani2016paradoxical}
Rangamani, P., Levy, M.G., Khan, S., Oster, G.: Paradoxical signaling regulates
  structural plasticity in dendritic spines.
\newblock Proceedings of the National Academy of Sciences \textbf{113}(36),
  E5298--E5307 (2016)

\bibitem{rao2002control}
Rao, C.V., Wolf, D.M., Arkin, A.P.: Control, exploitation and tolerance of
  intracellular noise.
\newblock Nature \textbf{420}(6912), 231--237 (2002)

\bibitem{sanchez2018oncogenic}
Sanchez-Vega, F., Mina, M., Armenia, J., Chatila, W.K., Luna, A., La, K.C.,
  Dimitriadoy, S., Liu, D.L., Kantheti, H.S., Saghafinia, S., et~al.: Oncogenic
  signaling pathways in the cancer genome atlas.
\newblock Cell \textbf{173}(2), 321--337 (2018)

\bibitem{sarma2012different}
Sarma, U., Ghosh, I.: Different designs of kinase-phosphatase interactions and
  phosphatase sequestration shapes the robustness and signal flow in the mapk
  cascade.
\newblock BMC systems biology \textbf{6}(1), 1--20 (2012)

\bibitem{sauro1994moiety}
Sauro, H.M.: Moiety-conserved cycles and metabolic control analysis: problems
  in sequestration and metabolic channelling.
\newblock BioSystems \textbf{33}(1), 55--67 (1994)

\bibitem{sauro2017control}
Sauro, H.M.: Control and regulation of pathways via negative feedback.
\newblock Journal of The Royal Society Interface \textbf{14}(127), 20160848
  (2017)

\bibitem{Sauro:MCABook}
Sauro, H.M.: Systems Biology: An Introduction to Metabolic Control Analysis.
\newblock Ambrosius Publishing (2018)

\bibitem{sauro2007mapk}
Sauro, H.M., Ingalls, B.: Mapk cascades as feedback amplifiers.
\newblock arXiv  (2007)

\bibitem{sauro2004quantitative}
Sauro, H.M., Kholodenko, B.N.: Quantitative analysis of signaling networks.
\newblock Progress in biophysics and molecular biology \textbf{86}(1), 5--43
  (2004)

\bibitem{sepulchre2012retroactive}
Sepulchre, J.A., Merajver, S.D., Ventura, A.C.: Retroactive signaling in short
  signaling pathways.
\newblock PloS one \textbf{7}(7), e40806 (2012)

\bibitem{shankaran2009rapid}
Shankaran, H., Ippolito, D.L., Chrisler, W.B., Resat, H., Bollinger, N.,
  Opresko, L.K., Wiley, H.S.: Rapid and sustained nuclear--cytoplasmic erk
  oscillations induced by epidermal growth factor.
\newblock Molecular systems biology \textbf{5}(1), 332 (2009)

\bibitem{shankaran2010oscillatory}
Shankaran, H., Wiley, H.S.: Oscillatory dynamics of the extracellular
  signal-regulated kinase pathway.
\newblock Current opinion in genetics \& development \textbf{20}(6), 650--655
  (2010)

\bibitem{shen2002network}
Shen-Orr, S.S., Milo, R., Mangan, S., Alon, U.: Network motifs in the
  transcriptional regulation network of escherichia coli.
\newblock Nature genetics \textbf{31}(1), 64--68 (2002)

\bibitem{shi2016conservation}
Shi, T., Niepel, M., McDermott, J.E., Gao, Y., Nicora, C.D., Chrisler, W.B.,
  Markillie, L.M., Petyuk, V.A., Smith, R.D., Rodland, K.D., et~al.:
  Conservation of protein abundance patterns reveals the regulatory
  architecture of the egfr-mapk pathway.
\newblock Science Signaling \textbf{9}(436), rs6--rs6 (2016)

\bibitem{shinar2007input}
Shinar, G., Milo, R., Mart{\'\i}nez, M.R., Alon, U.: Input--output robustness
  in simple bacterial signaling systems.
\newblock Proceedings of the National Academy of Sciences \textbf{104}(50),
  19931--19935 (2007)

\bibitem{small1990covalent}
Small, J.R., Fell, D.A.: Covalent modification and metabolic control analysis:
  Modification to the theorems and their application to metabolic systems
  containing covalently modifiable enzymes.
\newblock European journal of biochemistry \textbf{191}(2), 405--411 (1990)

\bibitem{sontag2009remarks}
Sontag, E.D.: Remarks on feedforward circuits, adaptation, and pulse memory.
\newblock IET Systems Biology \textbf{4}(1), 39--51 (2009)

\bibitem{spencer2010systemic}
Spencer, S.L., Albeck, J.G., Burke, J.M., Sorger, P.K., Gaudet, S., Kim, K.,
  Kim, D.H.: Systemic calibration of a cell signaling network model.
\newblock BMC Bioinformatics \textbf{11}(1), 1--14 (2010)

\bibitem{stadtman1977superiority}
Stadtman, E., Chock, P.: Superiority of interconvertible enzyme cascades in
  metabolic regulation: analysis of monocyclic systems.
\newblock Proceedings of the National Academy of Sciences \textbf{74}(7),
  2761--2765 (1977)

\bibitem{straube2017operating}
Straube, R.: Operating regimes of covalent modification cycles at high enzyme
  concentrations.
\newblock Journal of theoretical biology \textbf{431}, 39--48 (2017)

\bibitem{stroschein1999negative}
Stroschein, S.L., Wang, W., Zhou, S., Zhou, Q., Luo, K.: Negative feedback
  regulation of tgf-$\beta$ signaling by the snon oncoprotein.
\newblock Science \textbf{286}(5440), 771--774 (1999)

\bibitem{sturm2010mammalian}
Sturm, O.E., Orton, R., Grindlay, J., Birtwistle, M., Vyshemirsky, V., Gilbert,
  D., Calder, M., Pitt, A., Kholodenko, B., Kolch, W.: The mammalian mapk/erk
  pathway exhibits properties of a negative feedback amplifier.
\newblock Science signaling \textbf{3}(153), ra90--ra90 (2010)

\bibitem{suwanmajo2015mixed}
Suwanmajo, T., Krishnan, J.: Mixed mechanisms of multi-site phosphorylation.
\newblock Journal of The Royal Society Interface \textbf{12}(107), 20141405
  (2015)

\bibitem{takeda2012incoherent}
Takeda, K., Shao, D., Adler, M., Charest, P.G., Loomis, W.F., Levine, H.,
  Groisman, A., Rappel, W.J., Firtel, R.A.: Incoherent feedforward control
  governs adaptation of activated ras in a eukaryotic chemotaxis pathway.
\newblock Science signaling \textbf{5}(205), ra2--ra2 (2012)

\bibitem{ventura2010signaling}
Ventura, A.C., Jiang, P., Van~Wassenhove, L., Del~Vecchio, D., Merajver, S.D.,
  Ninfa, A.J.: Signaling properties of a covalent modification cycle are
  altered by a downstream target.
\newblock Proceedings of the National Academy of Sciences \textbf{107}(22),
  10032--10037 (2010)

\bibitem{ventura2008hidden}
Ventura, A.C., Sepulchre, J.A., Merajver, S.D.: A hidden feedback in signaling
  cascades is revealed.
\newblock PLoS Comput Biol \textbf{4}(3), e1000041 (2008)

\bibitem{walsh2014prospects}
Walsh, C.T., Wencewicz, T.A.: Prospects for new antibiotics: a
  molecule-centered perspective.
\newblock The Journal of antibiotics \textbf{67}(1), 7--22 (2014)

\bibitem{wynn2011kinase}
Wynn, M.L., Ventura, A.C., Sepulchre, J.A., Garc{\'\i}a, H.J., Merajver, S.D.:
  Kinase inhibitors can produce off-target effects and activate linked pathways
  by retroactivity.
\newblock BMC systems biology \textbf{5}(1), 1--15 (2011)

\bibitem{xiong2003positive}
Xiong, W., Ferrell, J.E.: A positive-feedback-based bistable ‘memory
  module’that governs a cell fate decision.
\newblock Nature \textbf{426}(6965), 460--465 (2003)

\bibitem{yi2000robust}
Yi, T.M., Huang, Y., Simon, M.I., Doyle, J.: Robust perfect adaptation in
  bacterial chemotaxis through integral feedback control.
\newblock Proceedings of the National Academy of Sciences \textbf{97}(9),
  4649--4653 (2000)

\bibitem{zumsande2010bifurcations}
Zumsande, M., Gross, T.: Bifurcations and chaos in the mapk signaling cascade.
\newblock Journal of theoretical biology \textbf{265}(3), 481--491 (2010)

\end{thebibliography}

\end{document}